\documentclass[aps,prd,a4paper,eqsecnum,%
superscriptaddress,nofootinbib,showpacs,%
twocolumn,showkeys,amsfonts,amssymb,amsmath]{revtex4}

\usepackage{epsfig}
\usepackage{amssymb,latexsym}
\usepackage{amsmath, amsthm}

\usepackage[all]{xy}

\newtheorem{theo}{Theorem}
\newtheorem{lemma}{Lemma}

\begin{document}

\title{\bf Dynamics of a self-gravitating shell of matter}

\author{Jerzy Kijowski}
\affiliation{Centrum Fizyki Teoretycznej PAN, Al.\ Lotnik\'ow
32/46, 02-668 Warszawa, Poland}

\author{Ewa Czuchry}
\affiliation{Katedra Metod Matematycznych Fizyki, ul. Ho\.{z}a 74,
00-682 Warszawa, Poland}

\begin{abstract}  Dynamics of a self-gravitating shell of matter is
derived from the Hilbert variational principle and then described
as an (infinite dimensional, constrained) Hamiltonian system. A
method used here enables us to define singular Riemann tensor of a
non-continuous connection {\em via} standard formulae of
differential geometry, with derivatives understood in the sense of
distributions. Bianchi identities for the singular curvature are
proved. They match the conservation laws for the singular
energy-momentum tensor of matter. Rosenfed-Belinfante and Noether
theorems are proved to be still valid in case of these singular
objects. Assumption about continuity of the four-dimensional
spacetime metric is widely discussed.
\end{abstract}



\maketitle

\section{Introduction}

In his seminal paper \cite{shell}, Werner Israel considered the
dynamics of a self-gravitating thin matter shell. The main purpose
of his theory was to find a simple model describing gravitational
collapse. In case of a realistic collapse, equations describing
evolution of matter and gravity are extremely difficult to handle.
Israel's idea was that many aspects of the collapse may be
investigated within a toy-model, which consists of a matter shell
and the surrounding gravitational field. The dynamics of such a
system reduces to a proper tailoring of the two different vacuum
solutions describing the two sides of the shell.

We are going to present a systematic derivation of the Israel
model from the variational principle and the construction of its
canonical (Hamiltonian) structure. In our approach the space-time
$M$ consists of two parts, tailored together along a hypersurface
$S$, which contains a moving matter shell. ``Tailoring'' means
that the induced metric $g_{ab}$ of $S$ is continuous. On the
other hand, the four-dimensional connection coefficients
$\Gamma^\lambda_{\mu\nu}$ may be discontinuous on $S$. It is
proved that the singular part of the Einstein curvature tensor
density of such a space-time contains derivatives of those
discontinuities and may be defined in the sense of distributions
as ${\cal G}^a{_b}={\bf G}^a{_b}\boldsymbol\delta_S$, where
$\boldsymbol\delta_S$ is a Dirac delta distribution concentrated
on $S$. As we show in the sequel, the following relation holds:
\begin{equation}
{\bf G}^a{_b}=[Q^a{_b}]\ ,
\end{equation}
where square brackets denote jump of the extrinsic curvature
across $S$, written in the ADM form. The singular Einstein tensor
density matches the singular (concentrated on $S$) energy-momentum
tensor density of the matter shell. Distributional Gauss-Codazzi
equations are then derived (and not {\em postulated}, as in the
Israel approach). They imply that both the Einstein tensor density
and the matter energy-momentum tensor density are conserved. We
prove that the conservation law  $\nabla_a {\bf G}^a{_b}=0$ can be
written in terms of the three-dimensional geometry of $S$.

The possibility of defining the singular Einstein tensor and
calculating its divergence (in the sense of distributions) {\em
via} the standard formulae of Riemannian geometry, simplifies
dramatically the calculational part of the theory. For this
purpose we must assume that the four-dimensional metric is
continuous across $S$. At this point our approach differs from the
technics used by many authors, including W.~Israel himself, who
always stressed that no assumptions about the continuity of the
four-metric, except for the continuity of the three-metric on the
surface $S$, are necessary. This is, of course, true. We observe,
however, that our ``additional condition'' on the continuity of
the entire four-dimensional metric, does not contain any
geometrical or physical condition imposed on configurations of the
physical fields considered by us, but is merely a gauge condition
imposed on coordinate systems used. Whenever an intrinsic
three-dimensional metric of $S$ is continuous, then also the
remaining four components of the four-dimensional metric may
become continuous after a suitable coordinate transformation. In
this new coordinate system we can use our technics based on the
distribution theory, but the dynamics derived this way is written
in terms of intrinsic, geometric relations, having sense in an
arbitrary system of reference. Hence, our derivation does not
depend upon particular gauge conditions which we use. Note that
even in a perfectly smooth, flat space-time, we can introduce a
coordinate system, in which only intrinsic three-dimensional
metric on a fixed hypersurface  $S=\{ x^3 =\mbox{const.} \}$ is
continuous, whereas the other four components of the
four-dimensional metric $g_{3 \mu}$ may be discontinuous. Using
these ``singular coordinates'' one can properly formulate
e.g.,~the initial value problem for the Maxwell field, but nobody
uses such a formulation for obvious reasons. Our additional
assumption on the continuity of space-time metric is motivated by
similar reasons: not to complicate things which already are
complicated enough. The technical flexibility we obtain this way,
i.e.,~the possibility of defining the singular geometric objects
living on $S$ {\em via} standard geometric formulae, where the
derivatives are understood in the sense of distribution, is the
main profit of our approach.

The structure of the paper is following. At the beginning we
derive Israel's model from variational principle, first in the
Lagrangian formulation, and then -- after performing an
appropriate Legendre transformation -- in the Hamiltonian picture.
Many technical elements of the construction presented here come
from the papers \cite{shell1} and \cite{K1}. A significant
simplification of the calculations is obtained using consequently
tensor densities instead of tensors for such quantities like the
matter energy-momentum and the Einstein curvature object (see also
\cite{JKC1}). Part of presented calculations was included in PhD
thesis of one of us \cite{ewa}.

\section{Proposed description of the Israel's model}\label{propon}

Consider a space-time consisting of two parts which are tailored
together along a hypersurface $S$, whose non-degenerate metric has
the signature $(-,+,+)$. Unlike in the original Israel's approach
(which was also used in \cite{shell1}), here we restrict ourselves
to coordinate systems for which all the components of the
space-time metric are {\em continuous}. As was already mentioned,
this condition does not limit the applicability of our formalism
and may be treated as merely a "gauge condition" imposed on the
coordinate system. It simplifies dramatically theoretical
description of the model. Further simplification is obtained by
using a coordinate system such that the hypersurface $S$ is given
by the equation $\{ x^3= {\rm const.}\}$. Metric derivatives along
$S$ (i.e.,~$\partial_a g_{\mu\nu}$, where $a=1,2,3\,$) are
continuous, whereas the transversal derivative (i.e.,~$\partial_3
g_{\mu\nu}$) may have jumps over $S$. We assume that the topology
of $S$ is of the type $S^2\times R^1$, i.e., it describes a
history of a matter concentrated on a two-dimensional surface with
the topology of the sphere $S^2$.

The shell divides the space-time into the internal and the
external part with respect to the world tube  $S$. In both parts
vacuum Einstein equations may be derived in a standard way from
the variational principle. Hence, the regular part of the Einstein
tensor must vanish everywhere outside and inside of $S$. Only its
singular part concentrated on $S$ is left.

The singular part of the Riemann tensor is proportional to the
(invariant) Dirac delta distribution $\boldsymbol\delta _S$
concentrated on $S$, because first derivatives of the metric (and,
whence, also the connection coefficients
${\Gamma}^{\lambda}_{\mu\nu}$) may be discontinuous across $S$. In
our particular coordinate system we have $\boldsymbol\delta
_S\equiv \boldsymbol\delta (x^3)$. As will bee seen in Section
\ref{Dynamics}, the resulting singular part of the Einstein tensor
will match the singular energy-momentum tensor describing matter
concentrated on $S$. This singular part may be obtained from the
standard formula for the Ricci tensor:
\begin{equation}\label{ric11}
  R_{\mu\nu}=\partial_\lambda{\Gamma}^{\lambda}_{\mu\nu}
  - \partial_{(\mu}
 {\Gamma}^{\lambda}_{\nu )
  \lambda}+{\Gamma}^{\lambda}_{\sigma\lambda}
 {\Gamma}^{\sigma}_{\mu\nu} - {\Gamma}^{\lambda}_{\mu\sigma}
 {\Gamma}^{\sigma}_{\nu\lambda} \ .
\end{equation}
A considerable simplification is obtained if we use the following
combination of Christoffel symbols:
\begin{equation}\label{A-def1}
A^{\lambda}_{\mu\nu} := {\Gamma}^{\lambda}_{\mu\nu} -
{\delta}^{\lambda}_{(\mu} {\Gamma}^{\kappa}_{\nu ) \kappa} \ .
\end{equation}
Then we have:
\begin{equation}
R_{\mu\nu}=\partial_\lambda
A^{\lambda}_{\mu\nu}-A^{\lambda}_{\mu\sigma}
A^{\sigma}_{\nu\lambda} + \frac 13 A^{\lambda}_{\mu\lambda}
A^{\sigma}_{\nu\sigma}.
\end{equation}
Because $A$ may have only discontinuities of the ``jump-type''
across $S$, the derivatives of $A$ along directions tangent to $S$
are thus finite and belong to the regular part of the Ricci
tensor. Hence, its singular part consist only of the transversal
derivatives:
\begin{equation}
{\rm sing}(R_{\mu\nu})=\partial_3
A^3_{\mu\nu}=\boldsymbol\delta(x^3)[A^3_{\mu\nu}] \ ,
\end{equation}
where the square brackets denote the jump of a specific quantity
across $S$. Hence, we have the following formula for the singular
part of the Einstein tensor density:
\begin{equation}\label{Einstein1}
  {\rm sing}\left( {{\cal G}^\mu}_\nu \right) :=
  \sqrt{|g|} \  {\rm sing}\left( { R^\mu}_\nu
  - \frac 12 R \right)  =
  \boldsymbol\delta(x^3) {{\bf G}^\mu}_\nu\ ,
\end{equation}
where
\begin{equation}\label{G-grube1}
  {{\bf G}^\mu}_\nu  :=  \sqrt{|g|} \left(\delta^\beta{_\nu} g^{\mu
  \alpha}  - \frac 12  \delta^\mu{_\nu} g^{ \alpha \beta} \right)
  [A^3_{\alpha\beta}]
\end{equation}
is a quantity living on $S$. Now, we are going to show that it is
actually a 3-dimensional tensor density on $S$. For this purpose
we first show that its components transversal to $S$ vanish and,
whence, its only non-trivial components are those tangential to
$S$. To prove that it transforms like a tensor density on $S$ let
us observe that ${{\cal G}^\mu}_\nu $ behaves like a 4-dimensional
density, which splits into the 3-dimensional density on $S$ and
the one-dimensional density along $x^3$. But the Dirac delta
$\delta(x^3)$ is already the density (and not a scalar!) on the
real axis $x^3$ which proves that the remaining object ${\bf G}$
behaves indeed like a 3-density. Hence, we have to prove the
following

\begin{lemma}

\begin{equation}
 \label{1G^3=0} {{\bf G}^\perp}_\nu \equiv 0 \ .
\end{equation}

\end{lemma}
\begin{proof}
On both sides of $S$ consider the following combination of the
connection coefficients:
\begin{equation}\label{def-qu-tylda0}
\tilde{Q}^\mu_{\ \nu} := \sqrt{|g|} \left( g^{\mu \alpha}
A^3_{\alpha\nu} - \frac 12 \delta^\mu_{\ \nu} g^{ \alpha \beta}
A^3_{\alpha\beta} \right) \ .
\end{equation}
It is useful to encode the entire information about the metric in
the following tensor density\footnote{In the so called affine
formulation of general relativity, proposed by one of us in 1978
(cite \cite{aff{}ine}), this quantity plays role of the momentum
canonically conjugate to connection  $\Gamma$.}
\begin{equation}\label{pi-kanon}
  {\pi}^{\mu\nu} := \frac 1{16 \pi}  \sqrt{|g|} \  g^{\mu\nu} \ .
\end{equation}
Hence, we have
\begin{equation}\label{def-qu-tylda1}
  \frac 1{16 \pi} \tilde{Q}^\mu_{\ \nu}
  = \pi^{\mu \alpha} A^3_{\alpha\nu} -
  \frac 12 \delta^\mu_{\ \nu} \pi^{ \alpha \beta} A^3_{\alpha\beta}
  \ ,
\end{equation}
Then the equation (\ref{G-grube1}) gives
\begin{equation}\label{G1}
  {\bf G}^\mu_{\ \nu} := [\tilde{Q}^\mu_{\ \nu} ] \ .
\end{equation}
Now, we use metricity condition for the connection $\Gamma$, which
is fulfilled on both sides of $S$ (indices $a,b=0,1,2 $ label the
coordinates on $S$):
\begin{eqnarray}
0 & \equiv & \nabla_a \pi^{33} = \partial_a \pi^{33} + 2
\pi^{3\mu} \Gamma^3_{\mu a} - \pi^{33} \Gamma^\mu_{a\mu}\nonumber \\
& &=
\partial_a \pi^{33}+ 2 \pi^{3\mu} A^3_{\mu a} \label{a-33} \\ 0
& \equiv & \nabla_a \pi^{3a} = \partial_a \pi^{3a} + \pi^{\mu a}
\Gamma^3_{\mu a} + \pi^{3\mu} \Gamma^a_{\mu a} - \pi^{3 a}
\Gamma^\mu_{a\mu} \nonumber
\\ & = & \partial_a \pi^{3a} + \pi^{ab}
A^3_{ab} - \pi^{33} A^3_{33} \ , \label{a-3a}
\end{eqnarray}
But the metric components  $\pi^{\mu\nu}$ and their derivatives
along $S$ are continuous across $S$. Hence, we obtain
\begin{align}\label{g31}
\frac 1{16 \pi} {\bf G}^3_{\ a} & =  \pi^{3 \mu} [A^3_{\mu a}] =-
\frac 12 [
\partial_a \pi^{33}] = 0 \ ,
\\
\frac 1{16 \pi} {\bf G}^3_{\ 3} & =  - \frac 12 \left( \pi^{ab}
[A^3_{ab}] - \pi^{33} [A^3_{33}] \right) =\frac 12 [
\partial_a \pi^{3a}] =  0 \ , \label{g32}
\end{align}
which ends the proof because in our coordinate system we have
${{\bf G}^\perp}_\nu ={{\bf G}^3}_\nu$.
\end{proof}

\section{Geometric interpretation of the singular curvature and
generalized Bianchi identities}\label{Bianchi}

Equation (\ref{G1}) expressing the singular part of Einstein
tensor in terms of jumps of quantities  $\tilde{Q}^\mu_{\ \nu}$
across $S$ is extremely useful in our derivation of the shell
dynamics. However, is not satisfactory from the geometric
viewpoint because quantities $\tilde{Q}^\mu_{\ \nu}$ do not have
any geometrical meaning on both sides of $S$, and only their jump
\eqref{G1} across $S$ does. Now, we are going to prove that we
obtain the same result replacing non-geometric object
$\tilde{Q}^\mu_{\ \nu}$ by a tensor density ${Q}^\mu_{\ \nu}$
which is, by definition, orthogonal to $S$ and whose restriction
${Q}^a_{\ b}$ to $S$ is equal to the external curvature of the
hipersurface $S$ in the ADM (\cite{ADM}) representation. In our
coordinate system $S = \{ x^3 = \mbox{\rm const.} \}$, external
curvature of $S$ is given by:
\begin{equation}\label{krzyw}
  L_{ab}:=-\frac{1}{\sqrt{g^{33}}}\ \Gamma^3_{ab}=
  -\frac{1}{\sqrt{g^{33}}}\ A^3_{ab} \ ,
\end{equation}
and its ADM version equals (cf. \cite{Misner}, \cite{K1}):
\begin{equation}\label{Q-pierwsz}
  {Q}^{ab}:=\sqrt{|\det g_{cd}|} \left(
  L\hat{g}^{ab}-L^{ab}\right)\ .
\end{equation}
Here, $\hat{g}^{ab}$ is a three-dimensional inverse of the induced
metric $g_{ab}$ on $S$ and $L^{ab}=\hat{g}^{ac} L_{cd} \,
\hat{g}^{db}$. It is easy to check that $\hat{g}^{ab}$ may be
calculated in terms of the four-dimensional inverse metric
$g^{\mu\nu}$ {\em via} the following formula (see \cite{Misner}):
\begin{equation}\label{odwrotna}
  \hat{g}^{ab} = {g}^{ab} - \frac {g^{3a} g^{3b}}{g^{33}} \ .
\end{equation}
In order to express coefficients $\tilde{Q}^\mu_{\ \nu}$ in terms
of the tensor density ${Q}^\mu_{\ \nu}$, observe that identities
(\ref{a-33}) and (\ref{a-3a}) can be solved algebraically with
respect to $A^3_{33}$ and $A^3_{3a}$. As a result, we have on both
sides of the hipersurface $S$ the following identities :
\begin{align}\label{MK1}
A^3_{33}&=\frac{1}{\pi^{33}}\left(
\partial_a \pi^{3a}+A^3_{ab} \pi^{ab}\right)\ ,
\\
A^3_{3a}&=-\frac{1}{2\pi^{33}}\left(
\partial_a\pi^{33}+2A^3_{ab}\pi^{3b}\right)\ .\label{MK2}
\end{align}
Hence, all the coefficients $A^3_{\mu\nu}$ can be expressed in
terms of $A^3_{ab}$, i.e.,, using (\ref{krzyw}) and
(\ref{Q-pierwsz}), in terms of ${Q}^{ab}$ and the metric. Using
again formula  (\ref{odwrotna}) we obtain:
\begin{equation}\label{tildeQ-Q}
  \tilde{Q}^a_{\ b} = {Q}^a_{\ b} + 16 \pi \left\{ - \frac 12 \frac
  {\pi^{3a}}{\pi^{33}} \pi^{33}_{\ \ ,b} + \frac 12 \delta^a_b
  \left( \frac{\pi^{3c}}{\pi^{33}} \pi^{33}_{\ \ ,c} -
  \pi^{3c}_{\ \ ,c} \right) \right\}\ .
\end{equation}
The last term is identical on both sides of  $S$, because the
metric components $\pi^{\mu\nu}$ are continuous. Hence, their jump
across $S$ vanishes and, due to (\ref{G1}), we obtain:
\begin{equation}\label{G1-n}
  {\bf G}^a_{\ b} := [{Q}^a_{\ b} ] \ ,
\end{equation}
whereas the transversal components ${\bf G}^\perp_{\ b}$ vanish.
Because the object  ${Q}^a_{\ b} $ is  a well defined tensor
density on both sides of $S$, (its definition does not depend upon
coordinate system used), hence the tensorial character of the
three-dimensional object ${\bf G}^a_{\ b}$ has been proved.

Now, we are going to show that the total Einstein tensor ${\cal
G}^\mu_{\ \nu}$ of our spacetime $M$ fulfills Bianchi identities.
Because regular part of ${\cal G}^\mu_{\ \nu}$ is discontinuous
across $S$ and, moreover, it contains also a Dirac-delta-like
singular part, these identities must be understood in a
distributional sense. To prove them we shall use Gauss-Codazzi
equations, relating transversal components ${\cal G}^{\perp}_{\
b}$ of the Einstein tensor density ${\cal G}^\mu_{\ \nu}$ to a
divergence of external curvature $Q$ on $S$:
\begin{equation}\label{G-C-niezd}
    {\cal G}^{\perp}_{\ b} +
    {\overline{\nabla} }_a {Q}^{a}_{\ b} \equiv 0
      \ .
\end{equation}
where ${\overline{\nabla} }$  denotes the intrinsic,
three-dimensional covariant derivative on $S$. But the transversal
component ${\cal G}^{\perp}_{\ b}$ is a well defined
three-dimensional object on $S$. In our coordinate system, adapted
to $S$ in such a way that the coordinate  $x^3$ is constant on
$S$, this quantity is simply equal to a ``third'' component:
${\cal G}^{\perp}_{\ b} = {\cal G}^{3}_{\ b}$. Taking a jump of
this equation across $S$ we obtain the following identity:
\begin{equation}\label{Bianchi-fund1-ndeg}
  [\mbox{\rm reg}({\cal G})^{\perp}_{\ b}] +
  {\overline{\nabla} }_a {\bf G}^{a}_{\ b}  \equiv 0
    \ .
\end{equation}
Now, we are going to show that the left hand side is exactly the
singular part of the Bianchi divergence: $\nabla_\mu {\cal
G}^{\mu}_{\ b} \equiv 0$. Indeed, the regular part of this
quantity vanishes on both sides of $S$ (a consequence of the
standard Bianchi identities), whereas its singular part is
proportional to $\boldsymbol\delta_S$. To prove this statement
observe that ${\cal G}^{3}_{\ b} = 0$ and, whence, no derivative
of the Dirac delta is produced when we apply the covariant
derivative $\nabla_3$ to the singular tensor density
(\ref{Einstein1}). What remains are thus the ``along $S$''
derivatives $\nabla_a$. As a result of this operation we obtain,
therefore, the quantity ${\overline{\nabla} }_a {\bf G}^{a}_{\ b}$
multiplied by $\boldsymbol\delta_S$. Another
$\boldsymbol\delta$-like term is obtained from the regular part
$\mbox{\rm reg}({\cal G})$ which is discontinuous across $S$.
Taking the derivative $\nabla_3 \mbox{\rm reg}({\cal G})^{3}_{\
b}$ of the regular part we obtain, therefore its jump of it across
$S$ multiplied by the Dirac delta: $[\mbox{\rm reg}({\cal
G})^{3}_{\ b}] \boldsymbol\delta_S $. Finally, the singular part
of the Bianchi identities is the sum of the above two expressions:
\begin{equation}\label{Bianchi-fund1-ndeg2}
    \nabla_\mu {\cal G}^{\mu}_{\ b} = \left(
  [\mbox{\rm reg}({\cal G})^{\perp}_{\ b}] +
  {\overline{\nabla} }_a {\bf G}^{a}_{\ b} \right)
  \boldsymbol\delta(x^3) \equiv 0
    \ ,
\end{equation}
where the last identity is just the Gauss-Codazzi equation
\eqref{Bianchi-fund1-ndeg}. Hence, we have shown that identities
$\nabla_\mu {\cal G}^{\mu}_{\ b} \equiv 0$ are also fulfilled for
space-times with a singular curvature.

\section{Dynamics of the ``shell + gravity'' system}\label{Dynamics}

Dynamical equations of the physical system composed of a matter
shell and the surrounding gravitational field will be derived from
the action principle $\delta {\cal A} = 0$, where
\begin{equation}\label{action}
  {\cal A} = {\cal A}_{\mbox{\tiny\rm grav}}^{\mbox{\tiny\rm reg}} +
  {\cal A}_{\mbox{\tiny\rm grav}}^{\mbox{\tiny\rm sing}} +
  {\cal A}_{\mbox{\tiny\rm matter}} \ ,
\end{equation}
is the sum of the gravitational action and the matter action.
Gravitational action, defined as the integral of the Hilbert
Lagrangian $L=\frac{1}{16\pi}\sqrt{g} R$, splits into the regular
${\cal A}_{\mbox{\tiny\rm grav}}^{\mbox{\tiny\rm reg}}$ and the
singular part ${\cal A}_{\mbox{\tiny\rm grav}}^{\mbox{\tiny\rm
sing}}$, according to the decomposition of the curvature
$R=\mbox{\rm reg}(R) + \mbox{\rm sing}(R)$ (a similar ``mixture''
of a ``bulk action'' and a singular ``body action'' concentrated
on a submanifold was recently used by C.~Barrab\`es and W.~Israel
-- see \cite{BI} -- to derive brane dynamics in general
relativity).

Using formulae (\ref{Einstein1})--(\ref{1G^3=0}), we express the
singular part of $R$ in terms of the singular part of the Einstein
tensor:
\begin{equation}\label{grav-sing}
\begin{split}
  16 \pi L_{\mbox{\tiny\rm grav}}^{\mbox{\tiny\rm sing}} = \sqrt{|g|} \
  \mbox{\rm sing}(R) = - \mbox{\rm sing}({\cal G}) \\=
  - {\bf G}^{\mu\nu} g_{\mu\nu} \boldsymbol\delta(x^3)
  = - {\bf G}^{ab} g_{ab} \boldsymbol\delta(x^3)\ .
\end{split}
\end{equation}
Hence the total action is the sum of three integrals:
\begin{equation}\label{dod}
    {\cal A} = \int_{D}
    L_{\mbox{\tiny\rm grav}}^{\mbox{\tiny\rm reg}} \
    + \int_{D}
    L_{\mbox{\tiny\rm grav}}^{\mbox{\tiny\rm sing}}  +
    \int_{D \cap S} L_{\mbox{\tiny\rm matter}} \ ,
\end{equation}
where $D$ is a spatially compact four-dimensional region with
boundary in $M$, which is possibly cut by a three-dimensional
surface $S$ (actually, because of the Dirac-delta factor, the
second term reduces to integration over ${D \cap S}$). Variation
is taken with respect to the spacetime metric tensor $g_{\mu\nu}$
and to the matter fields $z^K$ living on $S$. At the moment we do
not specify the nature of the matter fields. It is enough to
assume that they are geometric objects living on this surface and
that  matter Lagrangian $L_{\rm mat}$ is a scalar density on $S$,
which depends locally on the values of those fields, their
derivatives along $S$ and the metric of the surface $S$.

From the point of view of the two regular half-spacetimes (which
are tailored across $S$) the singular part of the action arises as
the sum of the boundary contributions from both the sides of the
shell. On the other hand, the action \eqref{dod} does not contain
any surface term at infinity. This is due to the techniques used
here (cf. \cite{K1}), where we first derive the field dynamics
within a spatially compact region $D$ and then shift its boundary
$\partial D$ to the space infinity. Of course, the boundary
manipulations at infinity are still necessary but in our approach
they arise as a Legendre transformation between different control
modes at the boundary $\partial D$ (see Section \ref{Legendre}).

There are many ways to calculate variation of the Hilbert
Lagrangian. Here, we use a method proposed by one of us (see
\cite{pieszy}). It is based on the following, simple observation:
\begin{eqnarray}
 \delta \left(  \frac 1{16 \pi} \sqrt{|g|} \ g^{\mu\nu} \
R_{\mu\nu} \right)\nonumber\\ =   - \frac 1{16 \pi} {\cal
G}^{\mu\nu} \delta g_{\mu\nu} &+& \frac 1{16 \pi} \sqrt{|g|} \
g^{\mu\nu} \delta R_{\mu\nu}
\nonumber \\
  =
 - \frac 1{16 \pi} {\cal G}^{\mu\nu} \delta g_{\mu\nu} &+&
 {\pi}^{\mu\nu} \delta R_{\mu\nu} \ .
 \label{deltaR1}
\end{eqnarray}
Expressing  $R_{\mu\nu}$ in (\ref{deltaR1}) by the connection
coefficients ${\Gamma}^{\lambda}_{\mu\nu}$ and their derivatives,
it is easy to show that the last term on the right hand side is a
complete divergence due to the following identity:
\begin{equation}\begin{split} \label{pidR}
  {\pi}^{\mu\nu} \delta R_{\mu\nu} \equiv
  \partial_\kappa \left( {\pi}_{\lambda}^{\ \mu\nu\kappa} \delta
  {\Gamma}^{\lambda}_{\mu\nu} \right)  =
  \left(\partial_\kappa  {\pi}_{\lambda}^{\ \mu\nu\kappa}\right)
   \delta  {\Gamma}^{\lambda}_{\mu\nu}\\ +
   {\pi}_{\lambda}^{\ \mu\nu\kappa} \delta
   {\Gamma}^{\lambda}_{\mu\nu , \kappa}
   \ ,
   \end{split}
\end{equation}
where, besides of the quantity (\ref{pi-kanon}), we have
introduced the following notation:
\begin{equation}\label{pi-41}
{\pi}_{\lambda}^{\ \mu\nu\kappa} := {\pi}^{\mu\nu}
\delta^\kappa{_\lambda} - {\pi}^{\kappa ( \nu} \delta^{\mu
)}{_\lambda} \ ,
\end{equation}
and
\begin{equation}\label{Gamma-4}
  {\Gamma}^{\lambda}_{\mu\nu , \kappa} := \partial_\kappa
  {\Gamma}^{\lambda}_{\mu\nu} \ .
\end{equation}
Proof of the  identity  (\ref{pidR}) is straightforward if one
uses the fact that  ${\Gamma}^{\lambda}_{\mu\nu}$ {\em are not}
independent quantities but  the connection coefficients,
i.e.,,~combinations of the metric components $g_{\mu\nu}$ and
their derivatives. This means that the covariant derivative
$\nabla {\pi}$ with respect to connection  $\Gamma$ vanishes
identically. Taking into account the fact that   $\pi$ is a tensor
density we have that:
\begin{equation}\label{delta-pi-41}
  \partial_\kappa {\pi}_{\lambda}^{\ \mu\nu\kappa} \equiv
  {\pi}_{\alpha}^{\ \mu\nu\kappa} {\Gamma}^{\alpha}_{\lambda\kappa}
  -{\pi}_{\lambda}^{\ \alpha\nu\kappa} {\Gamma}^{\mu}_{\alpha\kappa}
  -{\pi}_{\lambda}^{\ \mu\alpha\kappa} {\Gamma}^{\nu}_{\alpha\kappa}
  \ ,
\end{equation}
which implies identity (\ref{pidR}) in a simple way see
\cite{pieszy}).

Finally, variation of the regular part of the Hilbert Lagrangian
is following:
\begin{equation}\label{action-g1}
\begin{split}
  \delta \left(  \frac 1{16 \pi} \sqrt{|g|}
   \textrm{ reg}(R) \right) =
   - \frac 1{16 \pi}\textrm{ reg}( {\cal G})^{\mu\nu} \delta g_{\mu\nu}\\
   +\textrm{ reg}\left(
   \partial_\kappa \left( {\pi}_{\lambda}^{\ \mu\nu\kappa} \delta
   {\Gamma}^{\lambda}_{\mu\nu} \right) \right) \ .
   \end{split}
\end{equation}
We are going to show now that the same equation holds for the
singular part of $R$, i.e.,
\begin{eqnarray}
    \delta L_{\mbox{\tiny\rm grav}}^{\mbox{\tiny\rm sing}} &=&
  \delta \left(  \frac 1{16 \pi} \sqrt{|g|} \
   \textrm{ sing}(R) \right) \nonumber \\ &=&
   - \frac 1{16 \pi}\textrm{ sing}( {\cal G})^{\mu\nu} \delta g_{\mu\nu}
   +\textrm{ sing}\left(
   \partial_\kappa \left( {\pi}_{\lambda}^{\ \mu\nu\kappa} \delta
   {\Gamma}^{\lambda}_{\mu\nu} \right) \right) \ . \nonumber\\\label{action-g11}
\end{eqnarray}
\begin{proof}[Proof of (\ref{action-g11}):]
Calculate the singular part of $\partial_\kappa \left(
{\pi}_{\lambda}^{\ \mu\nu\kappa} \delta
   {\Gamma}^{\lambda}_{\mu\nu} \right)$.
 Because
all these quantities are invariant, geometric objects
($\delta\Gamma$ is a tensor!), we may calculate them in an
arbitrary coordinate system. Hence, we may use our adapted
coordinate system,  where coordinate $x^3$ is constant on $S$.
Taking into account the continuity of ${\pi}_{\lambda}^{\
\mu\nu\kappa}$ across $S$ we obtain
\begin{equation}\label{delta-S1}
\begin{split} \mbox{\rm sing}\left(
   \partial_\kappa \left( {\pi}_{\lambda}^{\ \mu\nu\kappa} \delta
   {\Gamma}^{\lambda}_{\mu\nu} \right) \right) =
  \boldsymbol\delta(x^3){\pi}_{\lambda}^{\ \mu\nu\perp} \delta
   [ {\Gamma}^{\lambda}_{\mu\nu} ] \\=
   \boldsymbol\delta(x^3){\pi}_{\lambda}^{\ \mu\nu 3} \delta
   [ {\Gamma}^{\lambda}_{\mu\nu} ] =\boldsymbol\delta(x^3)
   {\pi}^{\mu\nu} \delta
   [ A^{3}_{\mu\nu} ] \ .
   \end{split}
\end{equation}
From the definition (\ref{def-qu-tylda1}) of the object
$\tilde{Q}^{\mu\nu}$ it follows immediately that
\begin{equation}\label{A-przez-Q}
  {\pi}^{\mu\nu} \delta A^{3}_{\mu\nu} = - \frac 1{16 \pi}
  g_{\mu\nu} \delta {\widetilde  Q}^{\mu\nu}  \ .
\end{equation}
Using (\ref{G1}) we have that
\begin{equation}\label{tilde-Q1}
  [{\widetilde  Q}^{\mu\nu}]
  \boldsymbol\delta(x^3)
   =   \mbox{\rm sing}({\cal G})^{\mu\nu}
  \ .
\end{equation}
Putting these formulae together we obtain
\begin{equation}\label{Lgrav-sing1}
   \begin{split} \boldsymbol\delta(x^3)
   {\pi}_{\lambda}^{\ \mu\nu\perp} \delta
   [ {\Gamma}^{\lambda}_{\mu\nu} ] = -
   \frac 1{16 \pi} g_{\mu\nu} \delta \
   \mbox{\rm sing}({\cal G})^{\mu\nu} \\ =
   \delta L_{\mbox{\tiny\rm graw}}^{\mbox{\tiny\rm sing}}
   +\frac 1{16 \pi} \mbox{\rm sing}({\cal G})^{\mu\nu}
   \delta  g_{\mu\nu}
   \ ,
   \end{split}
\end{equation}
which ends the proof of  (\ref{action-g11}).
\end{proof}
Summing up the regular part (\ref{action-g1}) and the singular
part (\ref{action-g11}) we obtain variation of the whole
gravitational Lagrangian:
\begin{equation}\label{delta-A-grav-k1}
  \delta L_{\mbox{\tiny\rm graw}} =
  - \frac 1{16 \pi}
   {\cal G}^{\mu\nu} \delta g_{\mu\nu} +
   \partial_\kappa \left( {\pi}_{\lambda}^{\ \mu\nu\kappa} \delta
   {\Gamma}^{\lambda}_{\mu\nu} \right)\ ,
\end{equation}
which generalizes the corresponding formula from the paper
\cite{K1} to the case of spacetimes with a singular curvature.

\subsection{Matter Lagrangian and the singular energy-mo\-men\-tum}

Now, consider the matter part of the action. As was already
mentioned, we assume that all the dynamical properties of matter
fields are described by a Lagrangian density ${ L}_{\mbox{\tiny\rm
mat}}$, which is an invariant, three-dimensional scalar density on
$S$. It depends upon some matter fields $z^K$ living on $S$, their
first derivatives along $S$: ${z^K}_a:= \partial_a z^k$, and the
metric tensor  $g_{ab}$ on $S$:
\begin{equation}\label{Lmat1}
L_{\mbox{\tiny\rm mat}}=L_{\mbox{\tiny\rm mat}}(z^K;{z^K}_a;
g_{ab}) \ .
\end{equation}
Calculate the variation of the matter Lagrangian:
\begin{eqnarray}
  \delta L_{\mbox{\tiny\rm mat}} & = & \frac
  {\partial L_{\mbox{\tiny\rm mat}}}{\partial g_{ab}} \delta
  g_{ab} + \frac {\partial L_{\mbox{\tiny\rm mat}}}{\partial z^K}
  \delta z^K + \frac
  {\partial L_{\mbox{\tiny\rm mat}}}{\partial {z^K}_a}
  \partial_a \delta z^K \nonumber \\
  & = &  \frac 12 \tau^{ab} \delta g_{ab}
  + \left( \frac {\partial
  L_{\mbox{\tiny\rm mat}}}{\partial z^K} - \partial_a
  \frac
  {\partial L_{\mbox{\tiny\rm mat}}}{\partial {z^K}_a}
  \right) \delta z^K  \nonumber \\ &  & +
  \partial_a \left( {p_K}^a \delta z^K \right) \label{delta-L-matter1}
  \ ,
\end{eqnarray}
where we have introduced the following three-dimensional symmetric
energy-momentum tensor density  $\tau^{ab}$ on $S$:
\begin{equation}\label{sym}
   {\tau}^{ab}:=2 \frac{\partial L}{\partial g_{ab}} \ ,
\end{equation}
and the momenta $p_K{^a}$ canonically conjugate to material
variables $z^K$:
\begin{equation}\label{momentum-p1}
  {p_K}^a := \frac
  {\partial L_{\mbox{\tiny\rm mat}}}{\partial {z^K}_a} \ .
\end{equation}
Finally, we obtain the following formula for the variation of the
total (``matter + gravity'') Lagrangian:
\begin{align}\label{Lag-tot-var1}
  \delta L  = &  - \frac 1{16 \pi}
   \mbox{\rm reg}({\cal G})^{\mu\nu} \delta g_{\mu\nu} \nonumber\\
   &+
   \boldsymbol\delta(x^3)
   \left( \frac {\partial
  L_{\mbox{\tiny\rm mat}}}{\partial z^K} - \partial_a
  \frac
  {\partial L_{\mbox{\tiny\rm mat}}}{\partial {z^K}_a}
  \right) \delta z^K \nonumber \\
  & -  \boldsymbol\delta(x^3)
   \frac 1{16 \pi} \left( {\bf G}^{ab} - 8 \pi
     \tau^{ab}  \right) \delta g_{ab} \nonumber\\
  & +
   \partial_\kappa \left( {\pi}_{\lambda}^{\ \mu\nu\kappa} \delta
   {\Gamma}^{\lambda}_{\mu\nu} \right) +
   \boldsymbol\delta(x^3)
   \partial_a \left( {p_K}^a \delta z^K \right)
   \ .
\end{align}

In this Section we assume that both $\delta g_{\mu\nu}$ and
$\delta z^K$ vanish in a neighborhood of the boundary $\partial D$
of the space-time region $D$ (this assumption will be later
relaxed, when deriving Hamiltonian structure of the theory).
Hence, the last two boundary terms of the above formula vanish
when integrated over $D$. Vanishing of the variation $\delta {\cal
A} = 0$ with fixed boundary data implies, therefore, the
Euler-Lagrange equations  for the matter field $z^K$, together
with Einstein equations for gravitational field. Regular part of
Einstein equations:

\begin{equation}\label{E-reg}
    \mbox{\rm reg}({\cal G})^{\mu\nu} = 0
\end{equation}
is satisfied in the whole space-time, whereas the singular part
\begin{equation}\label{E-sing}
{\bf G}^{ab} = 8 \pi \tau^{ab} \ ,
\end{equation}
must be fulfilled on $S$. This way the Israel equations for the
shell dynamics have been derived from the variational principle.

Information encoded in the three-dimensional object $\tau^{ab}$
may be also expressed in terms of a four-dimensional
energy-momentum tensor  ${\cal T}^{\mu\nu} :=
\boldsymbol\delta(x^3) \tau^{\mu\nu}$, where, in agreement with
(\ref{1G^3=0}), the transversal components of $\tau$ vanish by
definition: $\tau^{\perp\nu} \equiv 0$ (in our adopted coordinate
system it simply means that $\tau^{3\nu} \equiv 0$, but this
condition in its previous form does not depend on the coordinate
system). Summing up the singular and the regular parts of the
gravitational field, we write the variation of the total
Lagrangian as
\begin{align}\label{Lag-1111}
  \delta L  = &  \frac 1{16 \pi}
   \left({\cal G}^{\mu\nu}  - 8 \pi
     {\cal T}^{\mu\nu}  \right)\delta g_{\mu\nu} \nonumber \\&+
   \boldsymbol\delta(x^3)
   \left( \frac {\partial
  L_{\mbox{\tiny\rm mat}}}{\partial z^K} - \partial_a
  \frac
  {\partial L_{\mbox{\tiny\rm mat}}}{\partial {z^K}_a}
  \right) \delta z^K \nonumber\\
  & +
   \partial_\kappa \left( {\pi}_{\lambda}^{\ \mu\nu\kappa} \delta
   {\Gamma}^{\lambda}_{\mu\nu} \right) +
   \boldsymbol\delta(x^3)
   \partial_a \left( {p_K}^a \delta z^K \right)
   \ ,
\end{align}
which is the starting point of our derivation of the dynamics of
the system. We stress that this equation is an {\em identity},
implied by the structure of the action \eqref{action}.

Field equations of the theory, i.e., Einstein equations for
gravitational field and the Euler-Lagrange equations for matter
fields, are equivalent to vanishing of the volume part of the
above variation (i.e., of the first two terms):
\begin{eqnarray}\label{Einstein-Eu-Lagr}
  {\cal G}^{\mu\nu}  & = & 8 \pi {\cal T}^{\mu\nu} \ ,
  \\ \label{matter-Eu-Lagr}
  \frac {\partial L_{\mbox{\tiny\rm mat}}}
  {\partial z^K} & = & \partial_a \frac
  {\partial L_{\mbox{\tiny\rm mat}}}{\partial {z^K}_a} \ .
\end{eqnarray}
This, in turn, is equivalent to the fact that for arbitrary
variations (non necessarily vanishing on the boundary!) of the
independent fields $\delta g_{\mu\nu}$ and $\delta z^K$, variation
of the Lagrangian reduces to the boundary part:
\begin{equation}\label{divergence1}
  \delta L =
   \partial_\kappa \left( {\pi}_{\lambda}^{\ \mu\nu\kappa} \delta
   {\Gamma}^{\lambda}_{\mu\nu} \right) +
   \boldsymbol\delta(x^3)
   \partial_a \left( {p_K}^a \delta z^K \right)
   \ .
\end{equation}
The whole dynamics of the system ``matter + gravity" is,
therefore, equivalent to the above equation. Similarly as in
equation (\ref{delta-S1}), we can use the definition of
${\pi}_{\lambda}^{\ \mu\nu\kappa}$ and express it in terms of
contravariant tensor density $\pi^{\mu\nu}$ obtaining
\begin{equation}
   {\pi}_{\lambda}^{\ \mu\nu \kappa} \delta
   {\Gamma}^{\lambda}_{\mu\nu}=\pi^{\mu\nu}\delta
   A^\kappa_{\mu\nu}\ .
\end{equation}
Hence field equations can be written in the following way:
\begin{equation}\label{divergence-A1}
  \delta L =
   \partial_\kappa \left( \pi^{\mu\nu}\delta
   A^\kappa_{\mu\nu} \right) +
   \boldsymbol\delta(x^3)
   \partial_a \left( {p_K}^a \delta z^K \right)
   \ .
\end{equation}

We complete this Section with the Noether Theorem for the
energy-momentum tensor \eqref{sym} which, due to Bianchi
identities \eqref{Bianchi-fund1-ndeg}, provides the necessary
consistency condition for the Einstein equations \eqref{E-sing}
or, equivalently, \eqref{Einstein-Eu-Lagr}. In fact, due to the
regular part \eqref{E-reg} of Einstein equations, Bianchi identity
\eqref{Bianchi-fund1-ndeg} reduces to: ${\overline{\nabla} }_a
{\bf G}^{a}_{\ b}  \equiv 0$. Hence, the following identity is
necessary and sufficient for the consistency of the singular part
\eqref{E-sing} of Einstein equations:

\begin{theo}

{\bf (Noether):} For any field configuration $(z^K(x))$ satisfying
the matter dynamics \eqref{matter-Eu-Lagr}, the energy-momentum
tensor \eqref{sym} carried by this configuration satisfies the
following identity:
\begin{equation}\label{Noether}
    {\overline{\nabla} }_a {\tau}^{a}_{\ b}  \equiv 0
\end{equation}

\end{theo}

The proof of the Noether identity is given in the next Section,
just after the Belinfante-Rosenfeld theorem.

\section{Hamiltonian structure of the theory}\label{Hamiltonian}

The above form of field equations is analogous to the Lagrangian
form of the dynamics in theoretical mechanics, which may be
written as follows:
\begin{equation}\label{mech-L}
  \delta L( q , {\dot q}) = \frac d{ dt} (p \delta q)
  \equiv {\dot p}\delta q + p \delta {\dot q} \ ,
\end{equation}
and contains relation between momenta and velocities:
\[
p = \frac {\partial L}{\partial {\dot q}}
 \ ,
\]
as well as Newton equations:
\[
{\dot p} = \frac {\partial L}{\partial {q}} \ .
\]
This formula is a starting point of derivation of the Hamiltonian
form of the dynamics. It is sufficient to perform Lagrange
transformation between  $p$ and ${\dot q}$, putting:
\[
p \delta {\dot q} = \delta (p {\dot q}) - {\dot q} \delta p \ ,
\]
and move the total derivative $\delta (p {\dot q})$ to the left
hand side of the equation (\ref{mech-L}). This way we obtain the
Hamiltonian formula:
\begin{equation}\label{Ham-mech}
  - \delta H(p,q) = {\dot p}\delta q - {\dot q} \delta p \ ,
\end{equation}
where we have put $H(p,q):= p{\dot q} - L$. This formula is
equivalent to the Hamiltonian form of the equations of motion:
\[
{\dot q} = \frac {\partial H}{\partial {p}} \ , \ \ \ \ \ {\dot p}
= - \frac {\partial H}{\partial {q}} \ .
\]

In order to derive the Hamiltonian formulation of the field
theory, we perform a similar Legendre transformation between time
derivatives of the fields and corresponding momenta. For this
purpose we have to fix a {(3+1)-decomposition} of the space-time
$M$. This way the theory becomes a Hamiltonian system, with the
space of Cauchy data on each of the three-dimensional surfaces
$\Sigma_t:=\{ t = \mbox{\rm const.}\}$ playing role of an
infinite-dimensional phase space. Unlike in case of the classical
mechanics, the dynamics of such a system is not uniquely defined,
unless we control also boundary data for the field in an
appropriate way.

In the present paper we consider the case of an asymptotically
flat space-time and assume that also the leaves $\Sigma_t$ of our
(3+1)-decomposition are asymptotically flat at infinity. To keep
control over 2-dimensional surface integrals at spatial infinity,
we first consider dynamics of our ``matter + gravity'' system in a
finite world tube ${\cal T}$, whose boundary carries a
non-degenerate metric of signature $(-,+,+)$. At the end of the
day we shall shift the boundary $\partial {\cal T}$ of the tube to
space-infinity. We assume that the tube contains the surface $S$
together with our matter travelling over it.

To simplify calculations we  choose coordinate system adapted to
this (3+1)-decomposition. This means that the time variable $t =
x^0$ is constant on three-dimensional surfaces of this foliation.
We assume that these surfaces are space-like. To obtain
Hamiltonian formulation of our theory we shall simply integrate
equation (\ref{divergence1}) (or, equivalently,
(\ref{divergence-A1})) over a finite piece $V$ of the Cauchy
surface ${\cal C} \subset M$ and then perform Legendre
transformation between time derivatives and the corresponding
momenta.

Denoting by $V := {\cal T} \cap {\cal C}$ the portion of the
Cauchy hypersurface ${\cal C}$ which is contained in the tube
${\cal T}$, we thus integrate (\ref{divergence-A1}) over the
finite volume $V \subset {\cal C}$ and keep surface integrals on
the boundary $\partial V$ of $V$. They will produce the ADM~mass
as the Hamiltonian of the total ``matter + gravity'' system  when
we pass to infinity with $\partial V = {\cal C} \cap
\partial {\cal T}$. Because our approach is geometric and does not
depend upon the choice of coordinate system, we may further
simplify our calculations using coordinate $x^3$ adapted to both
$S$ and to the boundary $\partial {\cal T}$ of the tube. We thus
assume that $x^3$ is constant on both these surfaces.

Integrating (\ref{divergence-A1}) over the volume $V$ we thus
obtain:
\begin{align}\label{generate-201}
  \delta\int_V  L =&\int_V \partial_\kappa \left(
  \pi^{\mu\nu}\delta
   A^\kappa_{\mu\nu} \right) +
   \int_V \boldsymbol\delta(x^3)
   \partial_a \left( {p_K}^a \delta z^K \right)\\
   =&\int_V\left( \pi^{\mu\nu}\delta
   A^0_{\mu\nu} \right)^\cdot + \int_{\partial V}
   \pi^{\mu\nu}\delta
   A^\perp_{\mu\nu} +\int_{V\cap S}
    \left( {p_K}^0 \delta z^K \right)^\cdot \ ,\label{generate-202}
\end{align}
where by ``dot'' we denote the time derivative. In this formula we
have skipped the two-dimensional divergencies which vanish when
integrated over surfaces $\partial V$ and $V\cap S$ (see fig. 1
below): \begin{center} \epsfig{file=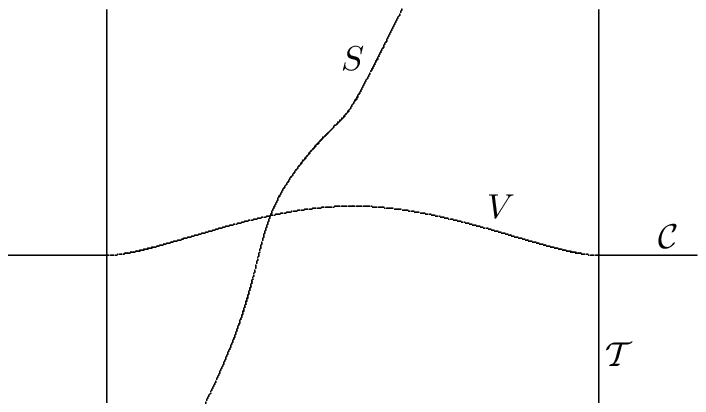,width=7truecm}
\end{center}

 To further simplify our formalism, we denote by $p_K :=
{p_K}^0$ the time-like component of the momentum canonically
conjugate to the field variable $z^K$.  Now we perform the
Legendre transformation in material variables:
\begin{equation}\label{legender-matter1}
  \left( p_K \delta z^K \right)^\cdot = \dot{p}_K \delta z^K -
  \dot{z}^K \delta p_K + \delta \left( p_K \dot{z}^K \right) \ .
\end{equation}
The last term, put on the left-hand side of (\ref{generate-201}),
meets the matter Lagrangian and produces the matter Hamiltonian
(with minus sign), according to formula:
\begin{equation}\label{T001}
  L_{\mbox{\tiny\rm mat}} - p_K \dot{z}^K =
  L_{\mbox{\tiny\rm mat}} - {p_K}^0 {z^K}_0 = - {T^0}_0  \ ,
\end{equation}
where the canonical energy-momentum tensor is defined as follows:
\[
{T^\mu}_\nu := {p_K}^\mu {z^K}_\nu - \delta^\mu_\nu \
L_{\mbox{\tiny\rm mat}}  \ .
\]

Even if singular, the above quantity satisfies the standard
Rosenfeld-Belinfante identity (cf. \cite{R-B}), which states that
the {\em canonical} energy-momentum tensor is equal (modulo a
minus sign, due to the convention used here) to the {\em
symmetric} energy-momentum tensor \eqref{sym}.

\begin{theo}

{\bf (Rosenfeld-Belinfante)}: Symmetric and canonical
energy-momentum tensors are, essentially, the same. More
precisely, the following identity holds:
\begin{equation}\label{R-B}
    {T^\mu}_\nu = - \tau^{\mu\lambda} \ g_{\lambda\nu} \ ,
\end{equation}
or, equivalently,
\begin{equation}\label{R-B-3D}
    {T^a}_b = - \tau^{ac} \ g_{cb} \ ,
\end{equation}
because both the transversal parts: ${T^\perp}_\nu$ and
$\tau^{\perp\nu}$ vanish identically from the definition.
\end{theo}

\begin{proof} We remember that $L_{\mbox{\tiny\rm
mat}}$ is a scalar density on $S$ and, therefore, may be written
as $L_{\mbox{\tiny\rm mat}} = \sqrt{\det g_{ab}}\ \Lambda$, where
$\Lambda$ is a scalar function, depending exclusively upon
quantities $(z^K;{z^K}_a; g_{ab})$. But the only way to produce a
scalar from the partial derivatives ${z^K}_a$ is to take the
following combination: $F^{KN}:={z^K}_a {z^N}_b \ \hat{g}^{ab}$.
We conclude that
\[
  L_{\mbox{\tiny\rm mat}} = \sqrt{\det g_{ab}}\ \Lambda (z^K , F^{KN})
 \ .
\]
This implies identity \eqref{R-B} in a straightforward way
(cf.~also \cite{shell1}).
\end{proof}

Now, we are ready for the proof of the Noether theorem
\eqref{Noether}:

\begin{proof}
 The invariant character of the matter Lagrangian
$L_{\mbox{\tiny\rm mat}}=L_{\mbox{\tiny\rm mat}}(z^K;{z^K}_a;
g_{ab})$ means that, for any vector field $X$ on $S$, dragging the
arguments $(z^K;{z^K}_a; g_{ab})$ along $X$ produces the same
effect on the Lagrangian that dragging it directly as a scalar
density. Choosing any coordinates $(x^a)$ on $S$ and choosing
$X=\partial_a$ we obtain, therefore, the following identity:
\begin{eqnarray}
  \partial_a L_{\mbox{\tiny\rm mat}} &=&
  \frac {\partial L_{\mbox{\tiny\rm mat}}}{\partial z^K} \
  {z^K}_a + \frac
  {\partial L_{\mbox{\tiny\rm mat}}}{\partial {z^K}_b}
  \ \partial_a {z^K}_{b} +  \frac
  {\partial L_{\mbox{\tiny\rm mat}}}{\partial g_{cd}}
  \ \partial_a g_{cd} \nonumber \\
   &=& \left( \frac {\partial L_{\mbox{\tiny\rm mat}}}{\partial z^K}
   - \partial_b \frac
  {\partial L_{\mbox{\tiny\rm mat}}}{\partial {z^K}_b} \right)
  \ {z^K}_a + \partial_b \left(
  \frac
  {\partial L_{\mbox{\tiny\rm mat}}}{\partial {z^K}_b}
  \ {z^K}_a \right)\nonumber\\&& + \frac 12 \tau^{cd}\ \partial_a g_{cd} \ ,
\end{eqnarray}
where we have used the symmetry of second derivatives: $\partial_a
{z^K}_{b} = \partial_b {z^K}_{a}$ and the definition \eqref{sym}
of the symmetric energy-momentum tensor $\tau$. Putting now the
last two terms on the left-hand side and using the
Rosenfeld-Belinfante theorem we obtain:
\begin{equation}\label{N-3}
     \partial_b {\tau^b}_a - \frac 12 \tau^{cd}\ \partial_a g_{cd}
    =  \left( \frac {\partial L_{\mbox{\tiny\rm mat}}}{\partial z^K}
   - \partial_b \frac
  {\partial L_{\mbox{\tiny\rm mat}}}{\partial {z^K}_b} \right)
  \ {z^K}_a
   \ .
\end{equation}
It may be easily checked that the left hand side is precisely the
covariant divergence ${\overline{\nabla} }_b {\tau}^{b}_{\ a}$ on
$S$ (remember that $\tau$ {\em is not} a tensor but the tensor
density!). Hence, we obtain a kinematic identity, fulfilled by
arbitrary field  configurations, not only those fulfilling field
equations:
\begin{equation}\label{N-4}
     {\overline{\nabla} }_b {\tau}^{b}_{\ a}
    =  \left( \frac {\partial L_{\mbox{\tiny\rm mat}}}{\partial z^K}
   - \partial_b \frac
  {\partial L_{\mbox{\tiny\rm mat}}}{\partial {z^K}_b} \right)
  \ {z^K}_a
   \ .
\end{equation}
This completes the proof.
\end{proof}

Now, integrating equation (\ref{T001}) over the two-dimensional
region $V\cap S$, we obtain the ``material'' part of the total
(``matter + gravity'') Hamiltonian:
\begin{equation}\label{Ham-matter}
  \int_{V\cap S} L_{\mbox{\tiny\rm mat}} - p_K \dot{z}^K
  = - \int_{V\cap S} {T^0}_0
  = \int_{V\cap S} {\tau^0}_0 = \int_{V} {{\cal T}^0}_0 \ ,
\end{equation}
where ${\cal T}^{\mu\nu} := \boldsymbol\delta(x^3) \tau^{\mu\nu}$.
Hence, the Legendre transformation in material degrees of freedom
gives us the following formula:
\begin{equation}\label{generate-201a}
\begin{split}
  \delta\int_V  \left({{\cal T}^0}_0 +  L_{\mbox{\tiny\rm graw}}
  \right) =\int_V\left( \pi^{\mu\nu}\delta
   A^0_{\mu\nu} \right)^\cdot + \int_{\partial V}
   \pi^{\mu\nu}\delta
   A^\perp_{\mu\nu} \\+\int_{V}
    \left(  \dot{\pi}_K\delta z^{K}-\dot{z}^{K}\delta
   \pi_K\right) \ ,
   \end{split}
\end{equation}
where we denote
\begin{equation}\label{pi-matter}
  \pi_K := p{_{K}} {\boldsymbol \delta}(x^3) \ .
\end{equation}
Here, the matter degrees of freedom are already described in the
Hamiltonian picture (with the matter Hamiltonian ``$-{{\cal
T}^0}_0$'' on the left hand side) and the gravitational degrees of
freedom still remain on the Lagrangian level.

\subsection{Legendre transformation in the gravitational degrees
of freedom}\label{Legendre}

To perform also Legendre transformation in gravitational degrees
of freedom --- analogous  to transformation
(\ref{legender-matter1}) in material degrees of freedom --- we
follow here a method proposed by one of us (see \cite{pieszy}). We
show in the Appendix that, after the transformation, formula
\eqref{generate-201a} assumes the following form:
\begin{eqnarray}
   &\delta &\,\int_V \left(
   {\cal T}^0{_0} -\frac{1}{8\pi}\,  {\cal G}^0{_0}\right)
   +
  \frac{1}{16\pi}\delta\int_{\partial
  V}\left( \mathcal{Q}^{00}g_{00}-\mathcal{Q}^{AB}g_{AB}\right)\nonumber
  \\&&
  \label{ham-pierw}  \\
  &=&
   \frac{1}{16\pi}\int_{V}\left(
  \dot{P}^{kl}\delta g_{kl}-\dot{g}_{kl}
  \delta P^{kl}\right)
  + \frac{1}{16\pi}
  \int_{\partial V}\left(  \dot{\lambda}\delta\alpha
  -\dot{\alpha}\delta\lambda\right) \nonumber \\
  & +& \int_{V}
    \left(  \dot{\pi}_K\delta z^{K}-\dot{z}^{K}\delta
   \pi_K\right)  - \frac{1}{16\pi}\int_{\partial V}g_{ab}
  \delta\mathcal{Q}^{ab} \label{control-bound}
  \ .
\end{eqnarray}
Here, we have introduced the following notation: $P^{kl}$ denotes
the external curvature of $\Sigma$ written in the ADM~form, i.e.,
given by the equations:
\begin{align}
P^{kl}:=  &  \sqrt{\det g_{mn}}\
(K{\tilde{g}}^{kl}-K^{kl})\label{Pklskrot} \ , \\
K_{kl}:=  &  -\frac{1}{\sqrt{|g^{00}|}}{\Gamma}_{kl}^{0}=-\frac{1}
{\sqrt{|g^{00}|}}A_{kl}^{0}\ , \nonumber
\end{align}
and ${\tilde{g}}^{kl}$ stands for three-dimensional contravariant
metric, invariant to the metric $g_{kl}$ induced on the Cauchy
surface $V$. Similarly, ${\cal Q}^{ab}$ denotes the external
curvature\footnote{We use the symbol ${\cal Q}$ for denoting
external curvature of the world tube  $\partial {\cal T}$ to
distinguish it from external curvature of the shell  $S$, which is
denoted by  $Q$.} of the tube $\partial {\cal T}$ written in the
ADM~form, i.e., three-dimensional tensor density given by
equations similar to (\ref{krzyw}) and (\ref{Q-pierwsz}):
\begin{align}
  {\cal Q}^{ab}:&=\sqrt{|\det g_{cd}|} \left(
  L\hat{g}^{ab}-L^{ab}\right)\ ,\label{qkl1skrot}  \\
  L_{ab}:&=-\frac{1}{\sqrt{g^{33}}}\Gamma^3_{ab}
  =-\frac{1}{\sqrt{g^{33}}}A^3_{ab}
  \ ,
\end{align}
and  $\hat{g}^{ab}$ is a three-dimensional contraindicant metric
on the tube $\partial {\cal T}$, invariant to the induced metric
$g_{ab}$. Moreover, $\lambda = \sqrt{\det g_{AB}}$ denotes the
two-dimensional volume form  on $\partial V$, whereas
\begin{equation}
  \alpha :={\rm ar\,sinh} \left(
  \frac{g^{30}}{\sqrt{| g^{00}g^{33}|}}\right) \ ,
\end{equation}
is the hyperbolic angle between the Cauchy surface $V$ and the
tube $\partial {\cal T}$.

The formula \eqref{ham-pierw} has been derived in the paper
\cite{K1} for a wide class of matter Lagrangian (including also
gauge fields), but only for models with the continuous matter
distribution. Now we have proved its  validity also in case of a
singular matter, concentrated on a two-dimensional shell $S$,
whose internal metric is non-degenerate and carries signature
$(-,+,+)$.

Observe that the first term on the left hand side of
\eqref{ham-pierw} vanishes identically due to Einstein
equations\footnote{The quantity $\left( {\cal G}^0{_0} - 8\pi
{\cal T}^0{_0} \right)$ in \eqref{ham-pierw} is often denoted by
$N{\cal H} + N^k {\cal H}_k$, where ${\cal H}$ and ${\cal H}_k$
are the scalar and the vector constraints respectively.} which
means that the volume part of the total ``gravity + matter''
energy vanishes identically. This does not mean that the energy
vanishes, because there is also another, surface contribution to
the Hamiltonian. Indeed, a detailed analysis (see \cite{pieszy})
shows that we are not allowed to control freely the tube external
curvature $\mathcal{Q}^{ab}$ in the last integral of
\eqref{control-bound} because of constraints which occur here. The
simplest way to overcome this difficulty is to perform another
Legendre transformation in the expression $g_{ab}
\delta\mathcal{Q}^{ab}  = g_{00}\delta\mathcal{Q}^{00} + 2
g_{0A}\delta\mathcal{Q}^{0A} + g_{AB}\delta\mathcal{Q}^{AB}$.
Namely, we write:
\begin{equation}\label{Legendre-g-q}
    g_{AB}\delta\mathcal{Q}^{AB} = \delta \left( g_{AB}\mathcal{Q}^{AB}
    \right) - \mathcal{Q}^{AB} \delta g_{AB} \ ,
\end{equation}
and put the complete derivative $\delta \left(
g_{AB}\mathcal{Q}^{AB} \right)$ on the left hand side of
\eqref{ham-pierw}. This way we obtain the ``quasi-local''
(i.e.,~assigned to the two-surface $\partial V$) Hamiltonian of
the system. Finally, we have:
\begin{eqnarray}\label{a}
  - \delta {\cal M}_{\partial V}
  &=&
   \frac{1}{16\pi}\int_{V}\left(
  \dot{P}^{kl}\delta g_{kl}-\dot{g}_{kl}
  \delta P^{kl}\right)\nonumber\\
&+& \frac{1}{16\pi}
  \int_{\partial V}\left(  \dot{\lambda}\delta\alpha
  -\dot{\alpha}\delta\lambda\right)  \\
  & +& \int_{V}
    \left(  \dot{\pi}_K\delta z^{K}-\dot{z}^{K}\delta
   \pi_K\right)  \nonumber\\
   &-& \frac{1}{16\pi}\int_{\partial V}
   g_{00}\delta\mathcal{Q}^{00} + 2
   g_{0A}\delta\mathcal{Q}^{0A} -
   \mathcal{Q}^{AB} \delta g_{AB}\label{b}
  \ ,\nonumber\\
\end{eqnarray}
where the quasi-local Hamiltonian (mass) assigned to the
two-dimensional surface ${\partial V}$ is defined as follows:
\begin{equation}\label{masa}
    {\cal M}_{\partial V}=E_0({\partial V}) - \frac{1}{16\pi}\int_{\partial
  V}\left( \mathcal{Q}^{00}g_{00}\right) \ ,
\end{equation}
and the additive constant $E_0({\partial V})$ is arbitrary. It
turns out (cf.~\cite{pieszy}) that it may be chosen in such a way
that the quasi-local mass vanishes for the flat Minkowski space
initial data\footnote{Actually, the quasi-local Hamiltonian
formula \eqref{a} needs some minor improvements, see
\cite{pieszy}, but it is sufficient for global purposes of the
present paper and this is why we do not discuss them here.}. The
final step of our derivation consists in shifting the tube ${\cal
T}$ to infinity. For this purpose we limit ourselves to the
asymptotically flat case and assume that the limiting case of $V$
is equal to an asymptotically flat Cauchy three-surface ${\cal
C}$, i.e.,~  $V \rightarrow {\cal C}$. It may be easily checked
that the limit of our quasi-local mass is equal to the ADM global
mass, i.e.,~that ${\cal M}_{\partial V} \rightarrow {\cal
M}_{\mbox{\tiny\rm ADM}}$. Moreover, all the surface terms on the
right hand side of our Hamiltonian formula \eqref{a}--\eqref{b}
vanish. This way we obtain the following global Hamiltonian
formula, fully analogous to the mechanical formula
\eqref{Ham-mech}, with the ADM mass ${\cal M}{\mbox{\tiny\rm
ADM}}$ playing role of the total hamiltonian of the ``matter +
gravity'' system:
\begin{align}
- \delta {\cal M}_{\mbox{\tiny\rm ADM}}  =  \frac 1{16 \pi}
\int_{\cal C} \left( {\dot P}^{kl} \delta g_{kl} - {\dot g}_{kl}
\delta P^{kl} \right) \nonumber\\+\int_{{\cal C}}\left(
\dot{\pi}_K\delta z^{K}-\dot{z}^{K}\delta \pi_K\right)
\label{dbarH-grav} \ .
\end{align}
The phase space of our system is thus described by the
gravitational variables $(P^{kl},g_{kl})$ on the Cauchy
three-surface ${\cal C}$, and by the material variables  $(\pi_K ,
z^K)$ concentrated on the two-surface $S \cap {\cal C}$,
cf.~\eqref{pi-matter}.

\section{Constraints}

As usual, Gauss-Codazzi equations imply constraints, which must be
fulfilled by the Cauchy data $(P^{kl} , g_{kl} , \pi_K , z^K)$ on
a three-surface ${\cal C}$. Outside of the shell these are
standard, vacuum constraints. In this Section we are going to
derive the complete description of constraints, valid not only for
the regular but also for the singular part of the data. We denote
by $\tilde{g}^{kl}$ the three-dimensional metric inverse to the
metric $g_{kl}$ and put $\gamma:=\sqrt{\det g_{kl}}$. By
$\stackrel{(3)}{R}$ denote the three-dimensional scalar curvature
of $g_{kl}$, $P:=P^{kl} g_{kl}$ and ``$|$'' is the
three-dimensional covariant derivative with respect to $g_{kl}$.

Outside of $S$ Gauss-Codazzi equations relate the components
${\cal G}^0{_\mu}$ of the Einstein tensor density with the Cauchy
data in the standard way. The spatial part of these constraints,
tangent to ${\cal C}_t$, reads as
\begin{equation}\label{GCww1}
{\cal G}^0{_l} =- P_l{^k}{_{|k}}\ ,
\end{equation}
and the time-like part -- normal to ${\cal C}_t$, as
\begin{equation}\label{GCws1}
 2{\cal G}^0{_\mu} n^{\mu} =
- \gamma\stackrel{(3)}{R}+\left(P^{kl}P_{kl}-\frac{1}{2}
P^2\right)\frac1\gamma \ ,
\end{equation}
where  $n$ denotes the future oriented, orthonormal vector to the
Cauchy surface ${\cal C}_t$:
\begin{equation}\label{w-ort}
n^\mu=- \frac {g^{0\mu}}{\sqrt{-g^{00}}}\ .
\end{equation}
Vacuum Einstein equations outside and inside of $S$ imply
vanishing of the regular part of ${\cal G}^0{_\mu}$. Hence, the
regular part of the vector constraint on  ${\cal C}$ reads:
\begin{equation}\label{v-reg}
    {\rm reg}\left( P_l{^k}{_{|k}}\right)=0 \ ,
\end{equation}
whereas the regular part of the scalar constraint reduces to:
\begin{equation}\label{s-reg}
   {\rm reg} \left(\gamma\stackrel{(3)}{R} -
   \left(P^{kl}P_{kl}-\frac{1}{2}
   P^2\right)\frac1\gamma  \right)= 0 \, .
\end{equation}

In the case of discontinuous matter, the above standard constraint
shall be  completed by their singular part, with support on the
intersection $S_{t} = {\cal C}_{t} \cap S$ in the following way.
Singular part of three dimensional derivatives of the ADM~momentum
$P_{kl}$ consists of derivatives in the direction of $x^3$:
\begin{equation}\label{v-sing}
   {\rm sing}(P_l{^k}_{|k}) ={\rm sing}(\partial_3 P_l{^3})=
   \boldsymbol\delta(x^3)[P_l{^3}]\ ,
\end{equation}
so the full Gauss-Codazzi equations (\ref{GCww1}) take the form
\begin{equation}\label{wwd1}
  {\cal G}^0{_l} =- {\rm sing}( P_l{^k}{_{|k}}) = -[P_l{^3}]
  \boldsymbol\delta(x^3)\ .
\end{equation}
Components of the ADM~momentum  $P^{kl}$ are regular, hence
singular part of the term $\left(P^{kl}P_{kl}-\frac{1}{2}
P^2\right)$ vanishes. Singular part of the three-dimensional
scalar curvature consists of derivatives in the direction of $x^3$
of the (three-dimensional) connection coefficients:
\[
\begin{split}
 {\rm sing}(\stackrel{(3)}{R})= {\rm sing} \left(
 \partial_3 \bigl( \Gamma^3_{kl}\tilde{g}^{kl}
 -\Gamma^m_{ml}\tilde{g}^{3l} \bigr)\right) \\=
 \boldsymbol\delta(x^3)\left[\Gamma^3_{kl}\tilde{g}^{kl}
 -\Gamma^m_{ml}\tilde{g}^{3l}\right]\, ,
\end{split}
\]
and expression in the square brackets may be reduced to the
following term
\begin{equation}
\begin{split} \gamma\left[\Gamma^3_{kl}\tilde{g}^{kl}
-\Gamma^m_{ml}\tilde{g}^{3l}\right]= -2\sqrt{\tilde{g}^{33}}
\left[\partial_3\biggl(\gamma\sqrt{\tilde{g}^{33}}\biggr)\right]
\\= -2\sqrt{\tilde{g}^{33}} \left[ \partial_k\left(\frac{\gamma
\tilde{g}^{3k}}{\sqrt{\tilde{g}^{33}}}\right)\right]  \, ,
\end{split}
\end{equation}
because derivatives tangent to  $S$ are continuous. But expression
in square brackets is equal to the external curvature scalar $k$
for the two-dimensional surface $S_t \subset {\cal C}_t$:
\begin{equation}
\gamma k= -\partial_k\left(\frac{\gamma
\tilde{g}^{3k}}{\sqrt{\tilde{g}^{33}}}\right) \, .
\end{equation}
This implies that
\[
{\rm sing}\left( \gamma \stackrel{(3)}{R}\right)= 2 \gamma
{\sqrt{\tilde{g}^{33}}}  [k] \boldsymbol\delta(x^3) = 2 [\lambda
k] \boldsymbol\delta(x^3) \ .
\]
Finally the total space-like Gauss-Codazzi equation  (\ref{GCws1})
takes the following form:
\begin{equation}\label{wsd1}\begin{split}
  2{\cal G}^0{_\mu} n^{\mu} = - {\rm sing}\left(
\gamma\stackrel{(3)}{R} - \left(P^{kl}P_{kl}-\frac{1}{2}
P^2\right)\frac{1}{\gamma} \right)\\
  = - 2 [\lambda k] \boldsymbol\delta(x^3) \, .
  \end{split}
\end{equation}
Equations (\ref{wwd1}) and (\ref{wsd1}) give a generalization (in
the sense of distributions) of the usual vacuum constraints
(vector and scalar respectively).

Now, we will show how the distributional matter located on $S_t$
determines the four surface quantities $[P^3_{\ k}]$ and $[\lambda
k]$, entering into the singular part of the constraints. The
tangent (to $S$) part of ${\cal G}^0{_\mu}$ splits into the
two-dimensional part tangent to $S_t$ and the transversal part.

The tangent to $S_{t}$ part of Einstein equations gives the
following:
\begin{equation}
{\cal G}^0{_k}=8\pi \boldsymbol\delta (x^3) \tau^0{_k} \ ,
\end{equation}
which, due to (\ref{GCww1}) and (\ref{wwd1}), implies the
following
 constraints:
\begin{equation}\label{PBt1}
\left[ P^3{_k}\right]=-8\pi\tau^0{_k}\ .
\end{equation}
The remaining null tangent part of Einstein equations reads:
$${\cal G}^0{_\mu}n^\mu= 8\pi{\cal T}^0{_\mu}n^\mu$$ where
$E:={\cal T}^0{_\mu}n^\mu={\tau}^0{_\mu}n^\mu {\boldsymbol
\delta}(x^3) = {\boldsymbol \delta}(x^3)\varepsilon$, describes
matter density on ${\cal C}$, and $\varepsilon =
{\tau}^0{_\mu}n^\mu$, is a surface energy density on $S$. Hence
$\textrm{sing}({\cal G}^0{_\mu}n^\mu)=8\pi\varepsilon$, and scalar
constraint takes the form
\begin{equation} \label{plk1}
  8\pi\varepsilon+ \left[\lambda k \right]=0\ .
\end{equation}
Finally, summing up the regular part of the constraints
(i.e.,~equations \eqref{v-reg} or \eqref{s-reg} respectively)
together with their singular parts (i.e.,~equations \eqref{v-sing}
or \eqref{wsd1} respectively) we may finally write down both
constraints in its distributional form:
\begin{eqnarray}
  P_l{^k}{_{|k}} &=& - 8\pi  \tau^0{_k} \ \boldsymbol \delta (x^3) \ ,\\
\hspace{-1cm}  \gamma\stackrel{(3)}{R} -
\left(P^{kl}P_{kl}-\frac{1}{2}
  P^2 \right) &=& - 16 \pi\varepsilon \ \boldsymbol \delta (x^3) \ ,
\end{eqnarray}
where the matter momentum $\tau^0{_k}$ and energy $\varepsilon$
must be expressed in terms of material variables $ (p_K , z^K)$
{\em via} the matter constitutive equations \eqref{sym}.

\section{Conclusions}
We have proved that the general scheme, used in \cite{pieszy} and
\cite{K1} to describe any continuous, self-gravitating matter, may
also be extended to the singular matter, concentrated on a
2-dimensional shell. The main result of this paper: the
Hamiltonian of the complete ``matter + gravity'' system is always
equal numerically to the ADM mass at infinity, similarly as in
continuous models. The above structure was used in \cite{jadwisin}
to derive the canonical formulation of a spherically symmetric
dust shell. Recently, it was proved that this result may be easily
extended far beyond the dust case (see \cite{Firenze}). We stress,
that the ADM mass generates the Hamiltonian evolution of the
system with respect to the asymptotic time variable at space
infinity, whereas the local redefinition of the Cauchy surface
${\cal C}_t:=\{ t = \mbox{\rm const.} \}$ surface, which does not
change it at infinity, is merely a gauge transformation. If we
want to use another time (i.~e.~the Minkowski time inside the
shell or the shell's proper time) there will be another
Hamiltonian generating the evolution in the new parametrization.
These issues were thoroughly discussed in \cite{shell1b}.

\section*{Appendix}

To prove formula \eqref{ham-pierw} {\em via} Legendre
transformation in gravitational degrees of freedom take metricity
conditions \eqref{MK1}--\eqref{MK2} (equivalent to equations
\eqref{a-33}--\eqref{a-3a}) for the connection $\Gamma$ on the
surface $\partial {\cal T} =\{ x^3 = \mbox{\rm const.} \}$ and
plug them into expression $\pi^{\mu\nu}\delta A^3_{\mu\nu} =
\pi^{33}\delta A^3_{33} + 2 \pi^{3a}\delta A^3_{3a} +
\pi^{ab}\delta A^3_{ab}$. A straightforward calculation leads to
the following result:
\begin{equation}\label{redukcja-31}
\begin{split}
  \pi^{\mu\nu}\delta A^\perp_{\mu\nu} =
    \pi^{\mu\nu}\delta A^3_{\mu\nu}=
    -\frac{1}{16\pi}g_{ab}\delta {\cal Q}^{ab}\\ +
    \partial_a\left(\pi^{33}\delta\left(
    \frac{\pi^{3a}}{\pi^{33}}\right)\right) \ ,
    \end{split}
\end{equation}
where ${\cal Q}^{ab}$ denotes the external curvature of the tube
$\partial {\cal T}$ written in the ADM~form, i.e., a
three-dimensional tensor density given by equations similar to
(\ref{krzyw}) and (\ref{Q-pierwsz}):
\begin{align}
  {\cal Q}^{ab}:&=\sqrt{|\det g_{cd}|} \left(
  L\hat{g}^{ab}-L^{ab}\right)\ ,\label{qkl1}  \\
  L_{ab}:&=-\frac{1}{\sqrt{g^{33}}}\Gamma^3_{ab}
  =-\frac{1}{\sqrt{g^{33}}}A^3_{ab}
  \ ,
\end{align}
and  $\hat{g}^{ab}$ is a three-dimensional contraindicant metric
on the tube $\partial {\cal T}$, inverse to the induced metric
$g_{ab}$.

Replacing now $x^3$ by $x^0$, we obtain analogous metricity
conditions on the surface $V = \{ x^0 = \mbox{\rm const.} \}$:
\begin{align}\label{MK3}
A^0_{00}&=\frac{1}{\pi^{00}}\left(
\partial_k \pi^{0k}+A^3_{kl} \pi^{kl}\right)\ ,
\\
A^0_{0k}&=-\frac{1}{2\pi^{00}}\left(
\partial_k\pi^{00}+2A^0_{kl}\pi^{0l}\right)\ .\label{MK4}
\end{align}
Plugging them into expression $\pi^{\mu\nu}\delta A^0_{\mu\nu} =
\pi^{00}\delta A^0_{00} + 2 \pi^{0k}\delta A^3_{0k} +
\pi^{kl}\delta A^0_{kl}$ we obtain immediately the following
identity:
\begin{equation}\label{redukcja-01}
    \pi^{\mu\nu}\delta A^0_{\mu\nu}=
    -\frac{1}{16\pi}g_{kl}\delta P^{kl} +
    \partial_k\left(\pi^{00}\delta\left(
    \frac{\pi^{0k}}{\pi^{00}}\right)\right) \ ,
\end{equation}
where $P^{kl}$ denotes the external curvature of $\Sigma$ written
in the ADM~form, i.e., given by the equations:
\begin{align}
P^{kl}:=  &  \sqrt{\det g_{mn}}\
(K{\tilde{g}}^{kl}-K^{kl})\label{Pkl1} \ , \\
K_{kl}:=  &  -\frac{1}{\sqrt{|g^{00}|}}{\Gamma}_{kl}^{0}=-\frac{1}
{\sqrt{|g^{00}|}}A_{kl}^{0}\ , \nonumber
\end{align}
and ${\tilde{g}}^{kl}$ stands for three-dimensional contravariant
metric, inverse to the metric $g_{kl}$ induced on the Cauchy
surface $V$.

Using these results and skipping the two-dimensional divergencies
which vanish after integration over $\partial V$, we may rewrite
the gravitational part of (\ref{generate-202}) in the following
way:
\begin{equation}\label{generate-grav1}
\begin{split}
   \int_V\left( \pi^{\mu\nu}\delta
   A^0_{\mu\nu} \right)^\cdot + \int_{\partial V}
   \pi^{\mu\nu}\delta
   A^\perp_{\mu\nu}  =
   -\frac{1}{16\pi}\int_V\left( g_{kl}\delta P^{kl} \right)^\cdot
   \\-\frac{1}{16\pi} \int_{\partial V} g_{ab}\delta {\cal Q}^{ab}
    +  \int_{\partial V} \left(\pi^{00}\delta\left(
    \frac{\pi^{03}}{\pi^{00}}\right) + \pi^{33}\delta\left(
    \frac{\pi^{30}}{\pi^{33}}\right) \right)^\cdot \ .
\end{split}\end{equation}
Next, we use the following, obvious identity

\begin{equation}
{\pi}^{00} \delta \left( \frac {{\pi}^{03}}{{\pi}^{00}} \right) +
{\pi}^{33} \delta \left( \frac {{\pi}^{30}}{{\pi}^{33}} \right) =
2 \sqrt{|{\pi}^{00}{\pi}^{33}|} \ \delta \frac
{{\pi}^{30}}{\sqrt{|{\pi}^{00}{\pi}^{33}|}} \ .
\end{equation}
Denote
\begin{equation}
q  :=  \frac {{\pi}^{30}}{\sqrt{|{\pi}^{00}{\pi}^{33}|}} = \frac
{g^{30}}{\sqrt{|g^{00}g^{33}|}} \ .
\end{equation}
We have
\begin{equation}
2 \sqrt{|{\pi}^{00}{\pi}^{33}|} = \frac 2{16 \pi} \sqrt{|g|}
\sqrt{|g^{00}g^{33}|} = \frac 1{8 \pi} \frac {\lambda}{\sqrt{1 +
q^2}} \ ,
\end{equation}
where $\lambda := \sqrt{\det g_{AB}}$. Hence, we obtain:
\begin{equation}\label{lambda-alpha1}
  \pi^{00}\delta\left(
    \frac{\pi^{03}}{\pi^{00}}\right) + \pi^{33}\delta\left(
    \frac{\pi^{30}}{\pi^{33}}\right) = \frac{1}{8\pi} \lambda
    \delta \alpha \ ,
\end{equation}
where $\alpha := {\rm arcsinh} (q)$ and, consequently,
\begin{align}
  &  \int_{V}\left(  \pi^{\mu\nu}
  \delta A_{\mu\nu}^{0}\right)  ^{\cdot}%
  +\int_{\partial V}\pi^{\mu\nu}\delta
  A_{\mu\nu}^{\perp} \label{generate-grav11}%
   =-\frac{1}{16\pi}\int_{V}\left(
  g_{kl}\delta P^{kl}\right)  ^{\cdot}\nonumber \\
&  -\frac{1}{16\pi}\int_{\partial
  V}g_{ab}\delta\mathcal{Q}^{ab}+\frac{1}{8\pi }\int_{\partial
  V}\left(  \lambda\delta\alpha\right)  ^{\cdot} \nonumber\ .
\end{align}
Now we perform Legendre transformation between time derivatives
and the corresponding canonical momenta. This transformation is
preformed both in volume:
\[
\left(  g_{kl}\delta P^{kl}\right)  ^{\cdot}=\left(
\dot{g}_{kl}\delta
P^{kl}-\dot{P}^{kl}\delta g_{kl}\right)  +\delta\left(  g_{kl}\dot{P}%
^{kl}\right)
\]
and on the boundary:
$(\lambda\delta\alpha)^{\cdot}=(\dot{\lambda}\delta
\alpha-\dot{\alpha}\delta\lambda)+\delta(\lambda\dot{\alpha})$.
The sum of the two total derivatives which arise here may be
easily calculated using the same arguments as in paper
\cite{pieszy}:
\begin{align}\label{pe-lambda1}
  -\frac{1}{16\pi}&\delta\int_{V}\left(  g_{kl}\dot P^{kl}\right)
  +\frac {1}{8\pi}\delta\int_{\partial
  V}\lambda\dot{\alpha}\nonumber\\
  &=\frac{1}{8\pi}\delta\int_V\sqrt{| g|}
  R^0{_0}+\frac{1}{16\pi}\delta\int_{\partial V}
  \left(  \mathcal{Q}^{AB}g_{AB}%
  -\mathcal{Q}^{00}g_{00}\right)\ .
\end{align}
Moving the first (volume) quantity to the left hand side of
formula (\ref{generate-201a}) and collecting it with the
gravitational part of the Lagrangian, we obtain:
\begin{align}
\frac{1}{16\pi}& \int_V L_{\mbox{\tiny\rm graw}} -
\frac{1}{8\pi}\int_V
\sqrt{|g|}R^0{_0}\nonumber\\&=\frac{1}{16\pi}\int_V
\sqrt{|g|}\left( R- 2R^0{_0}\right)= -\frac{1}{8\pi}\int_V {\cal
G}^0{_0}\ ,
\end{align}
which may be treated as the ``volume part of the
gravitational Hamiltonian''. It meets the ``matter Hamiltonian''
\eqref{Ham-matter}. Their sum (the ``volume part of the total
Hamiltonian'') vanishes identically as a consequence of Einstein
equations. This completes the proof of formula \eqref{ham-pierw}.


\end{document}